\documentclass[%
twocolumn,
 amsmath,amssymb,
pra,
10pt]{revtex4}

\usepackage{graphicx,color}
\usepackage{bm}
\usepackage{hyperref}

\definecolor{battleshipgrey}{rgb}{0.52, 0.52, 0.51}
\definecolor{cadet}{rgb}{0.33, 0.41, 0.47}
\definecolor{charcoal}{rgb}{0.21, 0.27, 0.31}

\hypersetup{
	colorlinks   = true, 
	urlcolor     = battleshipgrey, 
	linkcolor    = cadet, 
	citecolor   = charcoal 
}

\begin{document}

\preprint{APS/123-QED}

\title{Controlling Below-Threshold Nonsequential Double Ionization via Quantum Interference}

\author{A. S. Maxwell}
\email{andrew.maxwell.14@ucl.ac.uk}
\author{C. Figueira de Morisson Faria}%
\email{c.faria@ucl.ac.uk}
\affiliation{%
Department of Physics \& Astronomy, University College London \\Gower Street London  WC1E 6BT, United Kingdom
}%

\date{\today}

\begin{abstract}
We show through simulation that quantum interference in non-sequential double ionization can be used to control the recollision with subsequent ionization (RESI) mechanism. This includes the shape, localization and symmetry of RESI electron-momentum distributions, which may be shifted from a correlated to an anti-correlated distribution or vice versa, far below the direct ionization threshold intensity.  As a testing ground, we reproduce recent experimental results  by employing specific coherent superpositions of excitation channels. We examine two types of interference, from electron indistinguishability and intra-cycle events, and from different excitation channels.
\end{abstract}

\maketitle

Correlation and anti-correlation have been extensively studied in strong-field, laser-induced nonsequential double ionization (NSDI). In particular with the Cold Target Recoil Ion Momentum Spectroscopy (COLTRIMS) technique, information about the electron momenta has become experimentally accessible since the early 2000s. Several features in these distributions provide information about the type of interaction by which NSDI occur, and the physical mechanisms behind it. It is commonly accepted that NSDI results from the laser-induced inelastic recollision of an electron with its parent ion \cite{Corkum1993}. If the driving-field intensity is high enough, upon recollision the first active electron releases a second electron by electron-impact (EI) ionization. In contrast, for the so-called ``below-threshold" intensities, the kinetic energy transferred from the first electron to the core is not sufficient to free the second electron. Instead, the recolliding electron imparts only enough energy so that the second electron is excited, and then ionized with a time delay. This mechanism is known as recollision-exctiation  with subsequent ionization (RESI).

Throughout the years, the prevalent view is that, in EI, the two electrons will exhibit correlated momenta, as a consequence of their being released simultaneously. In contrast, for RESI, back-to-back emission will occur due to the time delay between recollision of the first electron and ionization of the second electron. This view has been backed using classical models, which have reproduced many of the key features encountered in experiments (for reviews see \cite{Faria2011}). Such models  also exhibit excellent agreement with the outcome of other methods, such as the full solution of the time-dependent Schr\"odinger equation \cite{Panfili2001}, or the strong-field approximation (SFA) \cite{FigueiradeMorissonFaria2004a,Faria2004,Jia2013}. The above-mentioned studies, performed for EI, suggested that quantum mechanical features such as interference will not survive integration over momentum components perpendicular to the laser-field polarization, which is the typical scenario in NSDI experiments. These conclusions were then extrapolated to RESI without much evidence.

Classical models have also successfully reproduced the anti-correlated behavior observed in early RESI experiments \cite{Eremina2003,Zeidler2005,Liu2008,Liu2010}. However, recent RESI experiments have found that electron-momentum distributions may occur in a variety of shapes, from cross-shaped distributions localized along the axis \cite{Bergues2012} to probability densities spread between all four quadrants of the plane spanned by the parallel electron momentum components \cite{Kubel2014, Sun2014}. These findings contradict the simple explanation that a time delay will lead to anti-correlated electron momenta.

A wealth of shapes has also been observed in many theoretical studies as diverse as the SFA and similar methods \cite{Shaaran2010,Shaaran2010a,Chen2010} to classical trajectory computations \cite{Emmanouilidou2009,Ye2010,Zhang2014}. If quantum interference is absent, the SFA retains four-fold symmetry for RESI distributions, in agreement with rigorous momentum constrains \cite{Shaaran2010a}.
Recently, however, SFA computations have shown that this symmetry can be broken, if the interference stemming from different excitation channels is incorporated \cite{Hao2014}. Therein, it has been argued that inter-channel interference is paramount for obtaining anti-correlated RESI distributions.

In this Letter, we show that one may obtain correlated, anti-correlated, cross- or ring-shaped RESI distributions in an SFA computation, by choosing appropriate coherent superpositions of excitation channels. We reproduce recent experimental results in which dramatic variations in the shapes of RESI distributions have been observed \cite{Kubel2014}. The features encountered are related to the interplay between two types of interference, involving (a) events which are displaced by half a cycle and those present due to the symmetry of indistinguishable electrons; and (b) different channels of excitation for the second electron. These effects individually break the four-fold symmetry of the momentum distributions, and may be used to manipulate the electron-electron correlation. 

In the SFA and in atomic units, the RESI transition amplitude related to the $c$-th excitation channel reads
\begin{eqnarray}
&&M^{(c)}(\mathbf{p}_{1},\mathbf{p}_{2})=\hspace*{-0.2cm}\int_{-\infty }^{\infty
}dt\int_{-\infty }^{t}dt^{^{\prime }}\int_{-\infty }^{t^{\prime
}}dt^{^{\prime \prime }}\int d^{3}k  \notag \\
&&\times V^{(c)}_{\mathbf{p}_{2}e}V^{(c)}_{\mathbf{p}_{1}e,\mathbf{k}g}V^{(c)}_{\mathbf{k}%
	g}\exp [iS^{(c)}(\mathbf{p}_{1},\mathbf{p}_{2},\mathbf{k},t,t^{\prime },t^{\prime
	\prime })],  \label{eq:Mp}
\end{eqnarray}
where the action
\begin{eqnarray}
&&S^{(c)}(\mathbf{p}_{1},\mathbf{p}_{2},\mathbf{k},t,t^{\prime },t^{\prime \prime
})=  \notag \\
&&\quad E_{\mathrm{1g}}t^{\prime \prime }+E_{\mathrm{2g}}t^{\prime
}+E^{(c)}_{\mathrm{2e}}(t-t^{\prime })-\int_{t^{\prime \prime }}^{t^{\prime }}%
\hspace{-0.1cm}\frac{[\mathbf{k}+\mathbf{A}(\tau )]^{2}}{2}d\tau  \notag \\
&&\quad -\int_{t^{\prime }}^{\infty }\hspace{-0.1cm}\frac{[\mathbf{p}_{1}+%
	\mathbf{A}(\tau )]^{2}}{2}d\tau -\int_{t}^{\infty }\hspace{-0.1cm}\frac{[%
	\mathbf{p}_{2}+\mathbf{A}(\tau )]^{2}}{2}d\tau  \label{eq:singlecS}
\end{eqnarray} describes the process in which an electron, initially at a bound state of energy $-E_{1g}$, leaves at $t^{\prime\prime}$, returns to the core at $t^{\prime}$ with intermediate momentum $\mathbf{k}$ and excites a second electron from a state with energy  $-E_{2g}$ to a state with energy  $-E^{(c)}_{2e}$. The first electron is released at $t^{\prime}$ with momentum $\mathbf{p}_1$, while the second electron is freed at a subsequent time $t$ with final momentum $\mathbf{p}_2$. The prefactors $V^{(c)}_{\mathbf{k}g}$, $V^{(c)}_{\bm{p}_1e,\mathbf{k}g}$ and $V^{(c)}_{\bm{p}_2e}$ are related to the ionization of the first electron, the recollision-excitation process and the tunnel ionization of the second electron, respectively. They contain all information about the interactions, which are chosen as long range, and electronic bound states \cite{Shaaran2010,Shaaran2010a}.
We compute the transition amplitude (\ref{eq:Mp}) using the steepest descent method, as described in \cite{Faria2002}.
We use the notation $p_{n\parallel}$ and $\mathbf{p}_{n\perp}$, $n=1,2$, for the momentum components parallel and perpendicular to the laser-field polarization, respectively. We employ a monochromatic field, which is a good approximation for longer pulses, and reasonable for short pulses if the carrier-envelope phase is averaged out.

 Due to momentum constraints attributed to the rescattering of the first electron and ionization of the second electron occurring most probably near field crossings and crests, respectively, the distributions will be located around $(p_{ 1\parallel},p_{ 2\parallel})=( \pm 2\sqrt{U_p},0)$. Considering the symmetry of displacement by half a cycle and particle exchange of the system leads to four ``events" located around $(-2\sqrt{U_p},0)$, $(2\sqrt{U_p},0)$, $(0,2\sqrt{U_p})$ and $(0,-2\sqrt{U_p})$. We label these event amplitudes as $M^{(c)}_l$, $M^{(c)}_r$, $M^{(c)}_u$ and $M^{(c)}_d$, given they occupy the left, upper, right and lower regions of the parallel momentum plane respectively. Quantum interference of events occurs mainly in the overlap of the above regions, which is located around the diagonals $p_{1\parallel}=\pm p_{2\parallel}$. 
 
\begin{figure}
\noindent\includegraphics[width=0.9\linewidth]{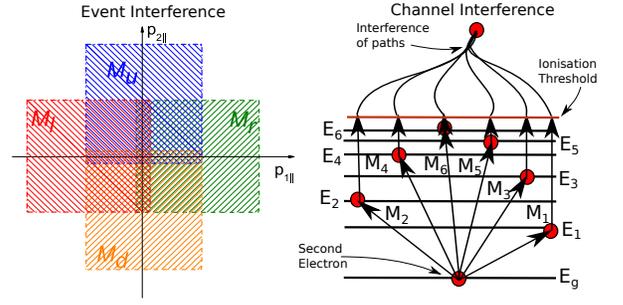}
\caption{\label{fig:interf}(Color online) Schematic representation of the event and inter-channel interference for RESI (panels (a) and (b), respectively). The shaded regions in panel (a) represent regions where $M^{(c)}_l$, $M^{(c)}_r$, $M^{(c)}_u$ and $M^{(c)}_d$ are substantial, and the arrows in panel (b) indicate the different excitation pathways for the second electron, according to Table \ref{table:channels}. }
\end{figure}

 For a single channel, the coherent sum of such amplitudes, integrated over the transverse momentum components reads
 \begin{align}
 	\Omega^{(c)}(p_{1\parallel},p_{2\parallel})= \int\int d^2 p_{1\perp}d^2 p_{2\perp}\left| M^{(c)}\right|^2,\label{Eq:Int}
 \end{align}
 with $M^{(c)}=M^{(c)}_l +M^{(c)}_r+ M^{(c)}_u+M^{(c)}_d$. If the events are summed incoherently, the amplitudes are replaced by probabilities, i.e., $|M^{(c)}_l|^2 +|M^{(c)}_r|^2+ |M^{(c)}_u|^2+|M^{(c)}_d|^2$ is employed instead.
If more than one channel is involved, Eq.~(\ref{Eq:Int}) is generalized to
\begin{align}
 \Omega(p_{1\parallel},p_{2\parallel})= \int\int d^2 p_{1\perp}d^2 p_{2\perp}\left|\sum_{c}M^{(c)}\right|^2. \label{Eq:Channels}
 \end{align}
Eq.~(\ref{Eq:Channels}) assumes that each excitation represents a path, which the second electron can take. Hence, the amplitudes corresponding to each path must be summed.  One may include different amplitudes or phases for each channel, which would model channel selection or account for phase effects. This yields a more general sum  $\int d^2p_{1 \perp}d^2p_{2 \perp}|\sum_c N_c e^{i \phi_c} M^{(c)}|^2$. In Fig.~\ref{fig:interf}, we provide a schematic representation of the event and inter-channel interference, for the excitation channels in Table \ref{table:channels}.
\begin{table}
	\begin{tabular}{c c c}
		\hline\hline
		Channel & Transition & $E_{2e}$ (a.u.)\\
		\hline
		1 & $3s3p^{6}$ ($3s \rightarrow 3p$ ) &0.52 \\
		2 & $3p^{5}3d$ ($3p \rightarrow 3d$) &0.41 \\
		3 & $3p^{5}4s$ ($3p \rightarrow 4s$) &0.4 \\
		4 & $3p^{5}4p$ ($3p \rightarrow 4p$) &0.31 \\
		5 & $3p^{5}4d$ ($3p \rightarrow 4d$) &0.18 \\
		6 & $3p^{5}5s$ ($3p \rightarrow 5s$) &0.19 \\
		\hline \hline
	\end{tabular}
	\caption{Excitation channels for $Ar^+$ employed in this work.
		 From left to right, the columns give the number associated with the channel, the electronic configurations for the sub-levels involved in the excitation and the absolute value $E_{2e}$ of the excited-state energy, respectively. The excitation pathway for
		the second electron is given in brackets. }
	\label{table:channels}
\vspace*{-0.35cm}
\end{table}
\begin{figure}
	\includegraphics[width=\linewidth]{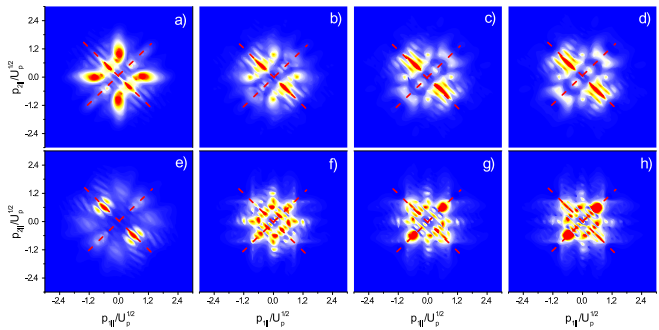}
	\caption{(Color online) RESI distributions for argon ($E_{1 g}= 0.58$, $E_{2 g}=1.02$ a.u.) computed using different coherent superpositions of the excitation channels in Table \ref{table:channels}. The phases and weights employed in these superpositions are provided in Table \ref{tab:Sup}, and have been chosen in such a way as to reproduce the experimental data from \cite{Kubel2014}. In the first row, panels a)-d), the intensity is $1\times 10^{14}$ W/cm$^{2}$ with $\omega=0.057$ a.u. and a ponderomotive energy of $U_p=0.22$ a.u. 
The intensities in the second row are as follows: e) $I=0.8\times 10^{14}$ W/cm$^{2}$ with $U_p=0.18$ a.u., f) and g) $I=0.8\times 10^{14}$ W/cm$^{2}$ with $U_p=0.26$ a.u. and h) $I=1.4\times 10^{14}$ W/cm$^{2}$ with $U_p=0.27$ a.u. The dashed lines in the figure indicate the diagonals $p_{1\parallel}=\pm p_{2\parallel}$.}
	\label{fig:Kubel}
\end{figure}

Experimental results from \cite{Kubel2014} of RESI in argon show that increasing pulse lengths from few to many cycle pulses causes a transition from cross-shaped to slightly anti-correlated, correlated or ring-shaped distributions.  Our results, displayed in Fig.~\ref{fig:Kubel}, exhibit many of these features. Specifically, we can model the transition from a cross to an anti-correlated distribution, as shown in the top row of Fig.~\ref{fig:Kubel}. Moreover, we can see in panels e) and f) that the distributions go from anti-correlated to correlated for increasing laser-field intensity. 

These effects are achieved by using coherent superpositions of the excitation channels given in Table \ref{table:channels}, using the relative weights and phases in Table.~\ref{tab:Sup}. The phases have been chosen to achieve an anti-correlated pattern in panels c). If a pattern is optimized to be anti-correlated at one driving field strength, changing this will cause it to flip to be correlated.  Hence, in panels f)-h) we obtain correlated patterns, in agreement with \cite{Kubel2014}. One should note that, despite this myriad of shapes, all intensities employed in this experiment are well within the below-threshold regime, as the maximal kinetic energy of the returning electron, $3.17U_p$, is significantly lower than the second ionization potential $E_{2g}$
\footnote{For the intensities in Fig.~\ref{fig:Kubel}, this energy varies from 0.55 a.u. to 0.86 a.u., i.e., between 54\% and 84\% of $E_{2g}$.}. Hence, RESI is the prevalent NSDI mechanism. Similar superpositions may be used to reproduce the results in \cite{Eremina2003,Liu2008,Liu2010}.
\begin{table}
	\begin{tabular}{c c c c c c c}
		\hline
		Panel (Fig.~\ref{fig:Kubel}) & 1 $(3p)$& 2 $(3d)$& 3 $(4s)$& 4 $(4p)$& 5 $(4d)$& 6 $(5s)$\\ \hline\hline
		a)& $1 e^{\frac{i\pi}{4}}$&	$1 e^0$&$4 e^{\frac{i7 \pi}{8}}$	 &$1 e^{\frac{i7 \pi}{8}}$&	 $0.5 e^{\frac{i\pi}{2}}$& $4 e^{\frac{i\pi}{2}}$\\
		b), e), f)& $1 e^{\frac{i\pi}{4}}$&	$1 e^0$&$2 e^{\frac{i7 \pi}{8}}$		&$1 e^{\frac{i7 \pi}{8}}$&	$0.6 e^{\frac{i\pi}{2}}$& $2 e^{\frac{i\pi}{2}}$\\
		c), g), h)& $1 e^{\frac{i\pi}{4}}$&	$1 e^0$&$1 e^{\frac{i7 \pi}{8}}$		&$1 e^{\frac{i7 \pi}{8}}$&	$0.7 e^{\frac{i\pi}{2}}$&	 $1 e^{\frac{i\pi}{2}}$\\
		d)& $1 e^{\frac{i\pi}{4}}$&	$1 e^0$&$0.5 e^{\frac{i7 \pi}{8}}$	&$2 e^{\frac{i7 \pi}{8}}$&	$0.8 e^{\frac{i\pi}{2}}$&	 $0.5 e^{\frac{i\pi}{2}}$\\
		\hline
	\end{tabular}
	\caption{Coherent superpositions employed in Fig.~\ref{fig:Kubel}. The letters in the first column indicate the panels in Fig.~\ref{fig:Kubel} for which a specific superposition have been employed. From the second to seventh column, the numbers in the first row indicate the excitation channel in Table \ref{table:channels}, and the excited state of the second electron is given in brackets. The numbers $N_ce^{i\phi_c}$ give the weight and the relative phase for each channel.}
	\label{tab:Sup}
\vspace*{-0.35cm}
\end{table}

\begin{figure*}
	\centering
	\includegraphics[width=\linewidth]{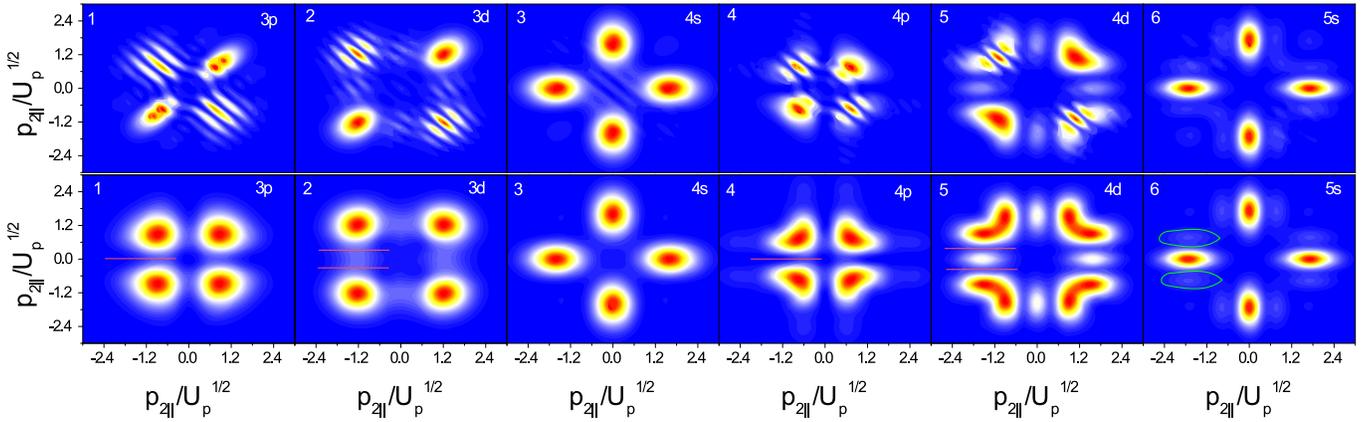}
	\caption{(Color online) One-channel RESI electron momentum distributions computed for argon with all prefactors included, for the parameters in Table \ref{table:channels}. The laser intensity and frequency are $I=4.56\times 10^{13}$ $\mathrm{W/cm}^2$ and $\omega=0.057$ a.u., respectively. Each panel has been labeled with the corresponding channel number (top left), as stated in Table \ref{table:channels}, and the excited state of the second electron (top right). The top and bottom rows display a coherent and incoherent sum of events, respectively. Angular and radial nodes from $V_{\mathbf{p}_2 e}$ have been marked on the bottom row with red lines and green circles, respectively. 
}
	\label{fig:fullprefactor}
\vspace*{-0.35cm}
\end{figure*}

\begin{figure}	
	\centering
	\includegraphics[width=0.9\linewidth]{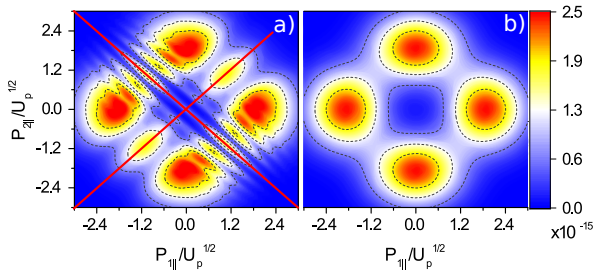}
	\caption{(Color online) One-channel coherent and incoherent sums $\Omega^{(c)}$ of all events, computed neglecting the prefactors for the same field parameters as in the previous figure [panel a) and b) respectively], and the RESI channel 1 in Table \ref{table:channels}.  The diagonals are indicated with the red lines in the figure. 
}
	\label{fig:FullMaps}
\vspace*{-0.35cm}
\end{figure}

The features observed in Figs.~\ref{fig:Kubel}(a) to (d) mark a change from a regime in which excitation to $s$ states is prevalent, to a scenario in which a coherent superposition of $p$ and $d$ states dominates. This statement can be inferred from Fig.~\ref{fig:fullprefactor}, which shows very different shapes for different channels of excitation. These differences stem from the prefactor $V_{\mathbf{p}_2 e}$ related to the ionization of the second electron. For $p$ and $d$ states, $V_{\mathbf{p}_2 e}$ has angular nodes, which prevent the distributions from being located along the axes. In contrast, for $s$ states only radial nodes are present, so that the electron-momentum distributions will be cross-shaped \cite{Shaaran2010,Maxwell2015}.
Furthermore, none of the coherent distributions in Figs.~\ref{fig:Kubel} and \ref{fig:fullprefactor} exhibit the fourfold symmetry obtained if incoherent sums are used. This symmetry breaking occurs already for a single channel, with fringes along and parallel to the two diagonals. These fringes may be understood if one considers a single-channel coherent sum (\ref{Eq:Int}) and its incoherent counterpart, neglecting the prefactors (see Fig.~\ref{fig:FullMaps}). They come from the interference of the amplitudes associated with electron indistinguishability. The diagonal maxima and minima satisfy the condition $p_{1\parallel}= p_{2\parallel}\pm |\delta|$, with $|\delta|\simeq \omega n \pi/(2\sqrt{U_p})$, where $n$ is even or odd, respectively. The anti-diagonal fringes are four times narrower. These fringes are obtained analytically by maximizing the integrand of Eq.~(\ref{Eq:Int}), and have been derived elsewhere \cite{Maxwell2015}.

In summary, we have shown that two types of quantum interference play an important role in reproducing results similar to those in experiments. Event interference will break the four-fold symmetry previously seen in the SFA, even when only considering a single channel of excitation.  Inter-channel interference can be used to maximize  anti-correlation, correlation,  and alter the shape of the electron-momentum distributions, and thus create all distributions found experimentally \cite{Eremina2003, Liu2008, Liu2010,Kubel2014}. 

This brings additional insight into many experimental studies \cite{Kubel2014,Liu2008,Eremina2003}, in which correlated distributions found for increasing driving-field intensity have been attributed to direct ionization, despite still being below the threshold intensity \cite{Liu2010}. Our results suggest that this could in fact be RESI, provided an appropriate coherent superposition of channels is chosen. Furthermore, recent experiments have shown by using a two-color field and changing the relative phase of the colors the momentum distributions for NSDI below the electron impact threshold can be manipulated from being anti-correlated to correlated \cite{Zhang2014b}. The phases of each channel and event have a strong dependence on the field as they derive from the action. Hence, it would not be unreasonable to assume this effect came from the interference we have described.

However, one may ask why very short pulses favor excitation to $s$ states, while longer pulses favor $p$ or $d$ channels. We have found that the relative s-state contributions decrease with the driving-field intensity, while those associated to the remaining channels remain relatively stable (not shown). For a few-cycle pulse, there are few dominant events near the peak of the pulse and other events near smaller crests. In contrast, the intensity across a long pulse will be more uniform. This may reduce the $s$ contributions for the same peak intensity. 

Finally, given the important role quantum interference has in RESI, one should consider the implications this has for classical models. We have shown that interference hugely increases the SFA's predictive power and can reproduce many of the features seen in experiment. Thus, classical-trajectories should be used with caution. Nonetheless, a large density of states may lead to a quasi-continuum, which could give rise to quasi-classical wave packets where interference would play less of a role. Our implementation of the SFA currently neglects broadening of states caused by the field. However, recent studies of the RESI phase-space dynamics have revealed a highly confined region that can be associated with trapping in an excited state \cite{Mauger2012}. This would justify using discrete bound states, and would render interference important. This provides a large scope for RESI to be used as an attosecond-imaging tool in order to probe and reconstruct the intermediate state of the second electron.

We thank J. Chen and X. Liu for discussions, B. Bergues for calling \cite{Kubel2014} to our attention, J. Dumont for his help with his complex Bessel function library \url{https://github.com/valandil/complex_bessel} and the UK EPSRC (grant EP/J019240/1) for financial support. The authors acknowledge the use of the UCL Legion High- Performance Computing Facility (Legion@UCL), and associated support services, in the completion of this work.

\end{document}